\documentclass{emulateapj}

\usepackage{amsmath}
\usepackage{amssymb}
\usepackage{graphicx}
\usepackage{grffile}
\usepackage[usenames,dvipsnames]{color}
\usepackage{bm}
\usepackage{natbib}
\usepackage{hyperref}
\usepackage{balance}

\newcommand{\HI}{{\text{H\MakeUppercase{\romannumeral 1}}} }

\newcommand{\Lya}{\ifmmode{{\rm Ly}\alpha}\else Ly$\alpha$\ \fi}
\newcommand{\cm}{\ifmmode{{\rm cm}}\else cm\fi}

\newcommand{\ergps}{\,{\rm erg}\,{\rm s}\ifmmode{}^{-1}\else ${}^{-1}$\fi}
\newcommand{\Mpch}{\,{\rm Mpc}\,\ifmmode h^{-1}\else $h^{-1}$\fi}
\newcommand{\kms}{\,\mathrm{km}\,\mathrm{s}^{-1}}
\newcommand{\dd}{\mathrm{d}}
\newcommand{\vek}[1]{\bm{#1}}
\def\gsim{~\rlap{$>$}{\lower 1.0ex\hbox{$\sim$}}}

\newcommand{\myfn}[1]{${}^{(\rm #1)}$}

\slugcomment{Sent to \apj\ 2015 May 27; accepted 2015 September 2; published 2015 October 15}

\begin{document}
\title{A systematic study of Lyman-Alpha transfer through outflowing shells: Model parameter estimation}

\author{M. Gronke, P. Bull and M. Dijkstra}
\affil{Institute of Theoretical Astrophysics, University of Oslo, Postboks 1029 Blindern, 0315 Oslo, Norway}
\email{maxbg@astro.uio.no}


\begin{abstract}
Outflows promote the escape of Lyman-$\alpha$ (Ly$\alpha$) photons from dusty interstellar media. The process of radiative transfer through interstellar outflows is often modelled by a spherically symmetric, geometrically thin shell of neutral gas that scatters photons emitted by a central Ly$\alpha$ source. Despite its simplified geometry, this `shell model' has been surprisingly successful at reproducing observed Ly$\alpha$ line shapes. In this paper we perform automated line fitting on a set of noisy simulated shell model spectra, in order to determine whether degeneracies exist between the different shell model parameters. While there are some significant degeneracies, we find that most parameters are accurately recovered, especially the HI column density ($N_{\rm HI}$) and outflow velocity ($v_{\rm exp}$). This work represents an important first step in determining how the shell model parameters relate to the actual physical properties of Ly$\alpha$ sources. To aid further exploration of the parameter space, we have made our simulated model spectra available through an interactive online tool.
\end{abstract}


\section{Introduction}
\label{sec:introduction}
Lyman-$\alpha$ (Ly$\alpha$) emission enables us to both find and identify galaxies out to the highest redshifts. Whether or not a distant galaxy has a \Lya emission line appears to be closely connected with the presence or absence of outflows \citep{1998A&A...334...11K,2008A&A...488..491A}. Partially coherent scattering of \Lya photons off the outflow provides an efficient way of shifting them into the wing of the line profile, where the galaxy is optically thin. Recent near-IR spectroscopic measurements have confirmed that the \Lya spectral line is systematically redshifted with respect to other non-resonant nebular lines \citep[e.g.][]{Steidel2010a,McLinden11,2012ApJ...745...33K,Schenker13,Erb2014,2014ApJ...791....3S,Shibuya14,Rivera-thorsen2014,Hashimoto15,Willott15}, which provides further support for the important role of (interstellar) outflows on the \Lya transfer process

The theoretical modelling of \Lya spectra has been the subject of many studies, both purely analytic \citep{Harrington73,Neufeld1990,Neufeld1991,Loeb1999,Dijkstra2006} and numerical \citep[e.g.][]{Loeb1999,Ahn00,Zheng2002,Iro06,Verhamme2006,Dijkstra2006,Hansen2005,Laursen2012,Duval2013,Behrens2014,Verhamme2014}. The latter use \Lya Monte-Carlo radiative transfer simulations, with the ultimate goal of understanding the \Lya spectra emerging from galaxies. The simulations can broadly be split into two categories:
\begin{itemize}
\item In the first class, the galactic environment is modelled as realistically as possible, mostly using hydrodynamical simulations \citep[e.g.][]{Iro06,Laursen07,Zheng2010,Verhamme12,Behrens2014a,Smith2014}. It is worth stressing that this is a challenging task as simulating the distribution and kinematics of neutral gas in feedback driven outflows -- which play key roles in the Ly$\alpha$ transfer process -- require a sub-pc spatial resolution \citep[compare e.g. figs.~10, 11 in][]{Fujita09}.

\item The second approach is to simplify the complex topography of galaxies enormously, and use an abstract geometrical setup instead. This approach is computationally cheap, which allows model \Lya spectra to be generated for a large set of parameters.
\end{itemize}

A particular example from the second group, which has been very successful in reproducing observed \Lya spectra, is the {\it shell model} \citep{Ahn2004ApJ...601L..25A,Verhamme2006,Schaerer2011}. This model consists of a central Ly$\alpha$ source surrounded by an outflowing thin, spherical shell of hydrogen and dust. The shell model has six model parameters: the equivalent width of the emitting source, $EW_i$; the intrinsic width of the \Lya spectrum, $\sigma_i$; the hydrogen column density, $N_{\HI}$; the dust optical depth, $\tau_d$; the temperature, $T$; and the outflow velocity, $v_{\rm exp}$.

Due to its simplicity, and the ability to reproduce a number of observed spectra, the parameters of the shell models have been frequently used to constrain galaxy properties \citep[e.g.][]{Verhamme2006,Verhamme2008,2007A&A...467...63T,2012ApJ...745...33K,2013ApJ...775...99C,2013ApJ...765..118W,Shibuya14,Martin2015,Hashimoto15}. 
These previous studies rely on either visual inspection and are, thus, affected by subjectivity, or, they use a discrete set of models from which they choose the best fitting spectrum based on $\chi^2$ values.
A systematic study of shell models and their physical relevance has yet to be performed, however. A number of open questions include the following.
\begin{enumerate}
\item Is there a unique mapping between shell model parameters and \Lya spectra, or can several different parameter combinations -- with different physical implications -- produce indistinguishable spectra? If the latter is true, two equally good fits to an observed spectrum could be found that have very different physical meanings, and so 
caution must be exercised when drawing conclusions from shell model parameter fits.
\item How uncertain are the parameters obtained from the model fitting procedure? The best-fitting shell model parameters to observed spectra are often quoted without estimates of uncertainties.
\end{enumerate}
The two questions are clearly related to one another: question (1) focuses on the intrinsic degeneracies between parameters, while question (2) concerns specific example spectra with some measurement uncertainty, and the effect this has on the accuracy with which the model parameters can be recovered.
Both must be understood if we are to use the observed spectral line shapes of \Lya to infer the physical properties of outflowing gas. This is especially important if one considers that, at high-$z$, \Lya is likely the only line that will allow us to put any constraints on outflows of atomic hydrogen gas.

The goal of this paper is to address these questions, and to develop a procedure to automatically fit shell model parameters to `realistic' (i.e. noisy) \Lya spectra. The paper is structured as follows. In Sec.~\ref{sec:method} we explain our method for constructing simulated \Lya spectra, and describe the fitting procedure we use. We present the results of applying this procedure to a range of simulated spectra in Sec.~\ref{sec:results}, and discuss the implications of our results in Sec.~\ref{sec:discussion}. We then conclude in Sec.~\ref{sec:conclusion}.

\section{Method}
\label{sec:method}
 We briefly describe the Ly$\alpha$ radiative transfer calculations in \S\ref{sec:lya-radi-transf}, and present our application to the shell model in \S\ref{sec:impl-shell-model}. Our fitting procedure is then described in \S\ref{sec:spectra-fitt-proc}.

In the following, $\log$ denotes the logarithm to base $10$, $\ln$ is used for the natural logarithm. The wavelength is given in units of velocity offset from line center, $v = c(\lambda/\lambda_0 - 1)$, where $\lambda_0\approx 1215.67$\AA \hspace{1mm} corresponds to the wavelength at line centre.

\subsection{\Lya radiative transfer}
\label{sec:lya-radi-transf}
We use the \Lya Monte Carlo (MC) radiative transfer code \texttt{tlac}, which was previously used in \cite{Gronke2014a}. We briefly summarize the main points of the algorithm here. For a more detailed review we refer the reader to \cite{Dijkstra2014_review}.

In a radiative transfer MC code, the photons are represented by individual particles (or rather, photon packets) whose paths of propagation in real and frequency space are tracked throughout the simulation domain. We repeat the following steps for each photon in the Monte-Carlo simulation:
\begin{enumerate}
\item A photon is emitted in direction $\vek{k}_i$, with intrinsic frequency $v_i$ drawn from some \Lya source emission pdf, $f(v)$.
\item A number $\tau$ is drawn from an exponential distribution, quantifying the distance $d$ the photon will travel. Here, $d$ is given by $\tau = \int_0^d (\sigma_{\HI}n_{\HI}(s) + \sigma_d n_{d}(s))\dd s$, where $n_X$ and $\sigma_X$ represent the number density and the cross section of dust (${}_d$) or hydrogen (${}_\HI$), respectively\footnote{Note that $\sigma_{\HI}$ depends strongly on frequency, which generally translates to a dependence of $\sigma_{\HI}$ on $s$. In addition, $\sigma_{\rm HI}$ is a function of the temperature $T$ \citep[see, e.g.,][for details]{Dijkstra2014_review}.}.
\item The position of the photon is updated to $\vek{p}\rightarrow \vek{p} + \vek{k} d$. At the new position, the photon is either scattered by hydrogen (probability of $n_{\HI}\sigma_{\HI} / (\sigma_{\HI}n_{\HI}(s) + \sigma_d n_{d})$), or absorbed by dust\footnote{For simplicity, we assume all-absorbing (i.e. non-scattering) dust throughout this work (that is, we set the dust albedo to $A=0$). We therefore have $\tau_{\rm a}\equiv (1-A)\tau_{d}=\tau_{d}$, where $\tau_{\rm a}$ [$\tau_{\rm d}$] denotes the optical depth to dust absorption [absorption \& scattering]. This assumption does not affect our predicted spectra \citep{Laursen09}, provided that results are `rescaled' as $\tau_{\rm d} \rightarrow \tau_{\rm d}/(1-A)$ when comparing to models with non-zero $A$.}. In the former case, a new direction is drawn from the proper phase function (i.e. the angular redistribution function) and a new frequency is drawn from the frequency redistribution function \citep[which depends on direction, see][]{DK12}.
\item Steps (2) and (3) are repeated until the photon escapes the simulation domain or is absorbed.
\end{enumerate}
If the photon escapes the simulation domain, its frequency and other properties are recorded. This procedure is repeated until the desired total number of photons is reached. Their combined frequencies then yield the simulated \Lya spectrum.
 
\begin{table}
  \caption{Simulation grid for shell model parameters}
  \begin{tabular}{@{}lll}
    \hline
    {\bf Symbol} & {\bf Values} & {\bf Units} \\
    \hline
    $v_{\rm exp}$\ \myfn{a} & $(0,2,5,8,10,15,20,30,40,\ldots,490)$ & $\kms$\\
    $\log N_{\rm HI}$\ \myfn{b}  & $(17.0, 17.2,\ldots,21.8)$ & $\log({\rm cm}^{-2})$\\
    $\log T$\ \myfn{c}  & $(3.0, 3.4, 3.8\ldots,5.6)$ & ${\log({\rm K})}$\\
    $\sigma_i$\ \myfn{d} & $[1,\,800]$ ~~~~~~~~~~~~~~~~~(continuous)& $\kms$\\
    $\tau_d$\ \myfn{e} & $[0,\,5]$ ~~~~~~~~~~~~~~~~~~~~(continuous)& --\\
    $EW_i$\ \myfn{f} & $[1,\,150]$ ~~~~~~~~~~~~~~~~~(continuous)& \AA \\
    \hline
  \end{tabular} 
 \myfn{a}expansion velocity,
 \myfn{b}hydrogen column density,
 \myfn{c}(effective) temperature,
 \myfn{d}width of intrinsic spectrum,
 \myfn{e}dust optical depth,
 \myfn{f}intrinsic equivalent width.
\\
  \label{tab:params}
\end{table}

\subsection{Implementation of the shell model}
\label{sec:impl-shell-model}
As discussed in Sec.~\ref{sec:introduction}, the shell model parameters consist of: (a) the source properties -- namely, the equivalent width of the emitting source $EW_i$, and the intrinsic width of the \Lya spectrum $\sigma_i$; and (b) the shell properties, i.e. its hydrogen column density $N_{\HI}$, its dust optical depth $\tau_d$ along a path that passes radially through the shell (the dust is confined to the neutral shell), temperature $T$, and the outflow velocity $v_{\rm exp}$.
In order to cover a particular domain of the parameter space with as few simulation runs as possible, we split these six parameters into two sets: the \textit{simulation} parameters $(v_{\rm exp},\,N_{\HI},\,T)$, and the \textit{post-processing} parameters $(EW_i,\,\sigma_i,\,\tau_d)$. The simulation parameters change the spectrum in a complex way. We therefore discretize these parameters on a grid and run our radiative transfer simulation at each combination of parameter values. This results in a grid of \Lya spectra with parameters presented in Table~\ref{tab:params}. The exact discretization was chosen so that the grid is fine enough to allow for an accurate likelihood analysis, i.e. so that the likelihood functions of our simulated spectra are sufficiently well sampled.
 
The post-processing parameters do not require explicit simulations, but can be modified {\it a posteriori}. This considerably reduces computational cost, and allows the parameters to be varied continuously (i.e. they are not restricted to discrete values on a grid).
Each photon has its own unique weight in our simulations which is given by
\begin{equation}
w = \frac{f_p(v_i)}{f_r(v_i)}e^{-\tau_d \hat N_{\HI}/N_{\HI}},
\label{eq:photon_weights}
\end{equation}
where $v_i$ is the velocity at which we inject the photon and $\hat N_{\HI}$ is the hydrogen column density actually encountered by the photon in the simulation\footnote{This column density $\hat N_{\HI} \geq N_{\HI}$ because scattering increases the total path a Ly$\alpha$ photons traverses before it escapes from the shell.}. Furthermore,  $f_r$  denotes the emission pdf that we used in our Monte-Carlo simulation (characterized by a Gaussian emission line with $\sigma=800\kms$ plus continuum, see below), while $f_p$ denotes the emission pdf that we wish to simulate in post-processing (which is also a Gaussian with standard deviation $\sigma_i$ plus continuum). 
This means $f_r$ and $f_p$ are of the form
\begin{align}
  f(v_i) = \begin{cases} \frac{1}{\eta + 1} \left(\frac{\eta}{\Delta v} + \mathcal{G}(v_i|\sigma_i)\right) &\text{for } |v_i|<\frac{\Delta v}{2}\\
0 & \text{otherwise}
  \end{cases}
\label{eq:intrinsic_pdf}
\end{align}
where $\mathcal{G}(v|\sigma)$ denotes a normalized Gaussian with standard deviation $\sigma$ centered at $v\!=\!0$, $\Delta v$ is the bandwidth of the spectrum, and $\eta\equiv \Delta v \lambda_{\Lya} / (c EW_i)$, where $\lambda_\Lya$ is the emission wavelength of Ly$\alpha$. We stress that \textit{each photon} is assigned an individual weight. This means, the post-processing parameters also affect the shape (and not only the normalization) of the resulting spectrum. We tested our method by comparing several post-processed spectra with `real' spectra where $\tau_d$, $\sigma_i$ and $EW_i$ were given as fixed input parameters in our Monte Carlo radiative transfer code.

Note that in this framework the escape fraction of \Lya photons $f_{\rm esc}$ can be calculated using
\begin{equation}
f_{\rm esc} = \frac{1}{\mathcal{N}}\sum\limits_{\rm photons}\frac{\mathcal{G}(v_i|\sigma_i)}{f_r(v_i)}\mathrm{e}^{-\tau \hat N_{\HI}/N_{\HI}}
\end{equation}
where $\mathcal{N} = \sum \mathcal{G}(v_i|\sigma_i)/f_r(v_i)$.

The quantity $f_{\rm esc}$ is of importance in both theoretical and observational studies \citep[e.g.][]{Blanc2011ApJ...736...31B,Hayes2011ApJ...730....8H,Dijkstra2013MNRAS.435.3333D} and can be used to further constrain the shell-model parameters (if measurements are available, see \S\ref{sec:param-uncert-degen}). Alternatively, the obtained $f_{\rm esc}$ (and its uncertainties) can be seen as an additional result of the shell-model fitting.

In order to minimize the uncertainty on the post-processed spectrum, the ratio $f_p(v_i)/f_r(v_i)$ should be as close to unity as possible across all frequency bins. When performing the grid simulations, we therefore chose a `typical' fiducial emission pdf, $f_r(v_i)$, with fixed values of $(\sigma_i,\,EW_i) = (800\kms,\,2.92\,{\rm \AA})$ for all simulations. The large $\sigma_i$ and small $EW_i$ ensure that a wide range of $v$ bins is initially populated with photons in the simulations, so $f_r > 0$ in any bin\footnote{A uniform emission pdf would achieve the same effect, but this would be inefficient, overpopulating the wings of the spectrum.}. The simulated value of dust optical depth is $\tau_d=0$ for minimizing computational time.

In our analysis, we allow the post-processing parameters to vary continuously between $(\sigma_i,\,\tau_d,\,EW_i) = ([1,\,800]\kms,\,[0,5],\,[1,\,150]\mathrm{\AA})$. These ranges were chosen to try and capture all physically-interesting scenarios, but are still somewhat arbitrary. They can be extended if necessary without requiring additional computations. 

The resulting dataset consists of $10,800$ spectra over a grid of the 3 simulated parameters, with $170,000$ photons per spectrum. The number of photons was chosen so that the approximate Poisson uncertainty (for a uniform spectrum) is $\lesssim 5\%$ on each bin; $\sqrt{N_{\rm bins} / 170,000} \sim 2.4\%$ ($N_{\rm bins}=100$). 
The spectra can be accessed through an interactive online tool\footnote{The online tool can be accessed under \url{http://maxgronke.wordpress.com/tools/tlac_web/} or\\ \url{http://bit.ly/man-alpha}. On the website, it is possible to \textit{(i)} plot up to four shell-model spectra, \textit{(ii)} upload and display own spectra, and, \textit{(iii)} download the plotted data.}.

\begin{figure*}
  \centering
  \plotone{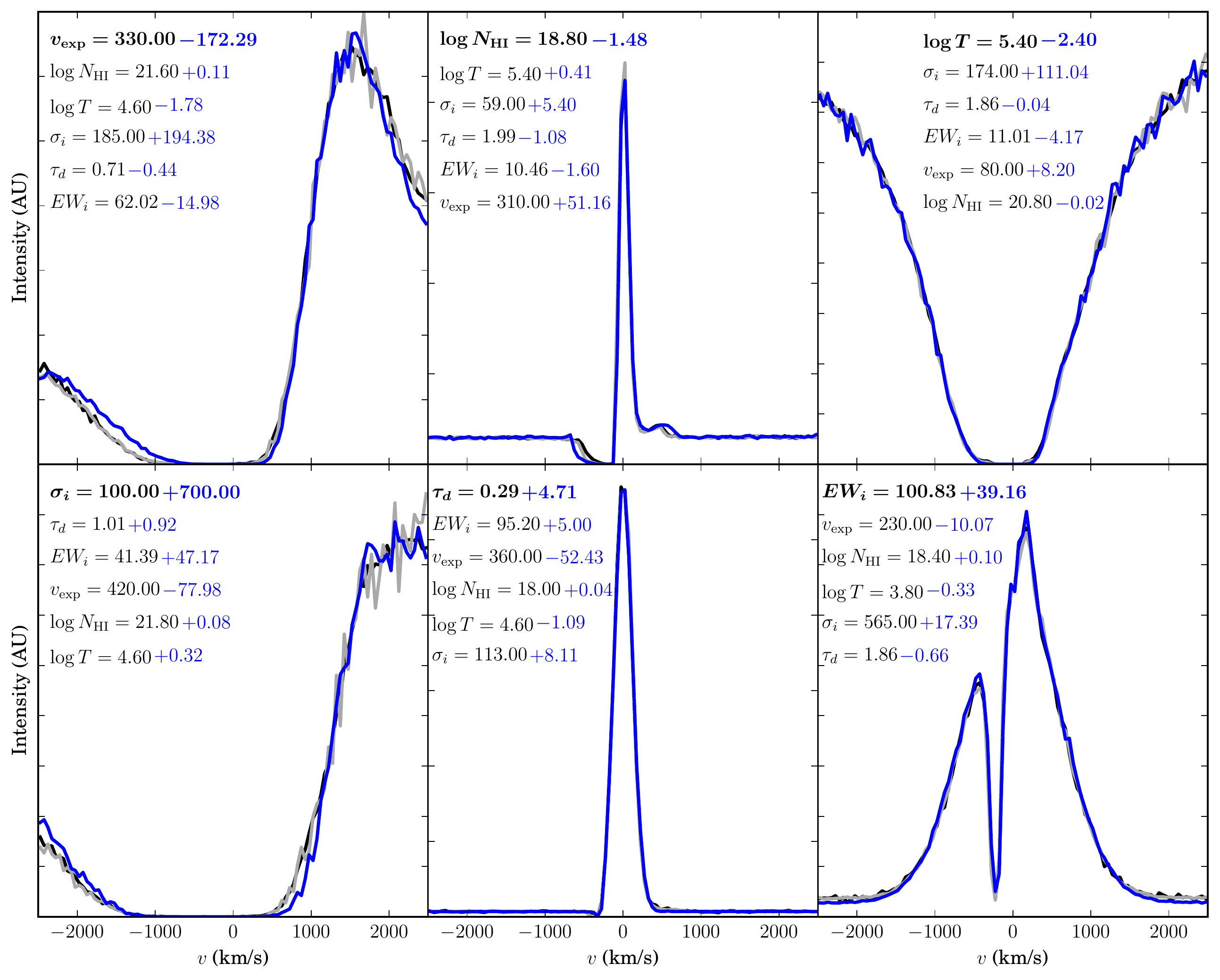}
  \caption{Examples of similar spectra produced by significantly different sets of shell model parameters. The `observed' spectrum is shown in gray, the spectrum obtained from the true shell-model parameters in black, and one of the best-fit models produced from 640 trials of the fitting procedure is shown in blue. The spectra were chosen to highlight extreme examples, one for each of the six model parameters (highlighted in bold in each panel). In each case, the fitted model is a better fit than the true model; see \S\ref{sec:intr-degen} for details. The parameters have the following units: $(v_{\rm exp},\,N_{\HI},\,T,\,\sigma_i,\,EW_i)=(\kms, \,\mathrm{cm}^{-2},\,\mathrm{K},\,\kms,\,{\rm \AA})$.}
  \label{fig:maxdist_multi}
\end{figure*}

\subsection{Simulated spectra and fitting procedure}
\label{sec:spectra-fitt-proc}

To test the effectiveness of spectrum fitting in recovering shell model parameters, we produced a fully-simulated dataset consisting of $50$ spectra for a range of randomly-chosen input parameters.\footnote{These were drawn from a uniform distribution with the bounds given in Table~\ref{tab:params}, except for $\tau_d$, which was drawn from $[0,\,2]$ to save on computational resources.} The simulated spectra were found directly by using the method described in \S\ref{sec:lya-radi-transf}, i.e., no post-processing was used, yielding the `true' (input) spectrum, $F(v_i; \vek{\theta}_0)$, for each set of input parameters $\vek{\theta}_0$, and then adding noise, $n_i$, to each spectral bin $i$, resulting in an `observed' spectrum
\begin{equation}
d_i = F(v_i; \vek{\theta}_0) + n_i.
\end{equation}
While in reality the noise level in each bin will be a complicated function of instrumental characteristics, observing conditions, and the properties of the data reduction pipeline, for simplicity (and to avoid tying our results to a particular observational configuration) we chose $n_i$ to be Gaussian, with zero mean, constant variance, and no correlations between frequency bins. The frequency binning was chosen as constant intervals of width $\Delta v_i = 50\kms$ over the range\footnote{While we expect this range to be sufficient for most observed spectra (i.e. wide enough to sample the \Lya line and the adjacent continuum), it is possible to alter this range if needed since our method is not dependent on this choice.} $v=[-2500,\,2500]\kms$, with a simple tophat bandpass and no smoothing. This corresponds to a relatively high (but plausible) resolving power of $R = c/\Delta v_i\sim 6000$, which lies within the capabilities of, e.g., the \textit{IRCS} instrument on the \textit{Subaru} telescope\footnote{\url{http://subarutelescope.org/Observing/Instruments}.}. We discuss the impact of changing the binning in \S\ref{sec:snr}.

Following our assumption of uncorrelated Gaussian noise, the appropriate goodness-of-fit statistic for the spectra is 
\begin{equation}
\Delta\chi^2 = \sum\limits_i \left ( \frac{d_i - F(v_i; \vek{\theta})}{\hat{\sigma}_i} \right )^2,
\label{eq:logp}
\end{equation}
where the sum is taken over the frequency bins. This defines the likelihood $\mathcal{L}(\vek{\theta} | d) \propto \mathrm{e}^{-\Delta\chi^2 / 2}$. We systematically explore the likelihood surfaces of our simulated data using Markov Chain Monte Carlo (MCMC) and nonlinear optimization (fitting) methods in the next section. Since we have assumed that the noise rms (i.e., the statistical weight) $\hat{\sigma}_i = \hat{\sigma}(v_i) = {\rm const.}$, and is independent of model parameters $\vek{\theta}$, the statistic
\begin{equation}
\Delta^2 = \sum\limits_i \left( d_i - F(v_i; \vek{\theta}) \right)^2
\label{eq:Deltasq}
\end{equation}
can be used interchangeably with $\Delta\chi^2$ by optimization algorithms, a fact that we will also use in Sec.~\ref{sec:results}.

Note that there is also a `theoretical uncertainty' associated with the model spectrum $F(v; \vek{\theta})$. Because each spectrum is calculated through a Monte Carlo radiative transfer procedure with a finite number of photons, there is a Poisson error on the value of $F$ in each bin, determined by the number of photons that ended up in that bin (and their respective weights). Since the number of photons per bin also varies as a function of the input parameters, the theoretical uncertainty will vary across the parameter space in a potentially complicated way. This uncertainty has been mitigated as much as possible by our use of a large number of photons to calculate each model on the grid of simulations, as described in \S\ref{sec:impl-shell-model}. This helps to ensure that the assumed observational uncertainty, $\hat{\sigma}_i$, dominates the Poisson uncertainty in each bin. Under these conditions, the theoretical uncertainty can safely be ignored, and Eq.~(\ref{eq:logp}) remains valid.

\begin{figure*}
  \centering
  \plotone{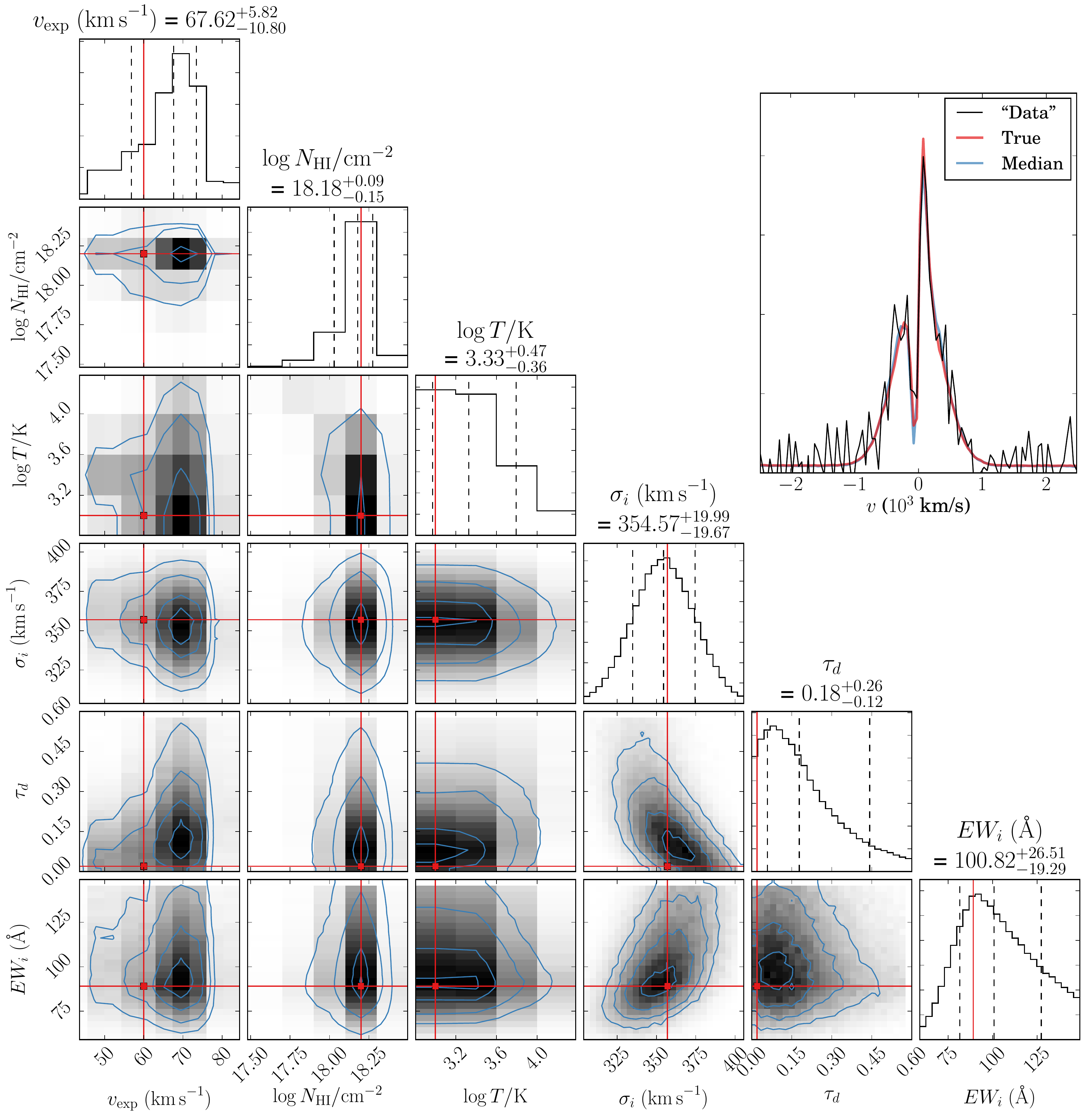}
  \caption{Example analysis of an asymmetric double-peaked profile. {\it Upper right}: Simulated data (black solid line) with its associated 68\% CL observational uncertainty (gray shaded region), and the spectra for the true (input) and median MCMC-estimated parameters (light red and light blue, respectively). {\it Main plot}: The 1D histograms and 2D contour plots show the one- and two-dimensional marginal posterior distributions from the MCMC chains respectively. The red solid line and red marker show the true (input) model parameters. In the histograms, the dashed lines mark the $16$th, $50$th and $84$th percentiles; these numbers are also stated on top of each column. In the contour plots, the blue lines mark the $(2,\,1.5,\,1,\,0.5)\,\sigma$ contours, and the gray shading gives the posterior density. Recall that the parameters $\log N_{\rm HI}$, $\log T$ and $v_{\rm exp}$ are discrete (visible through the blocky structure of the contours).}
  \label{fig:mcmc1}
\end{figure*}

\begin{figure*}
  \centering
  \plotone{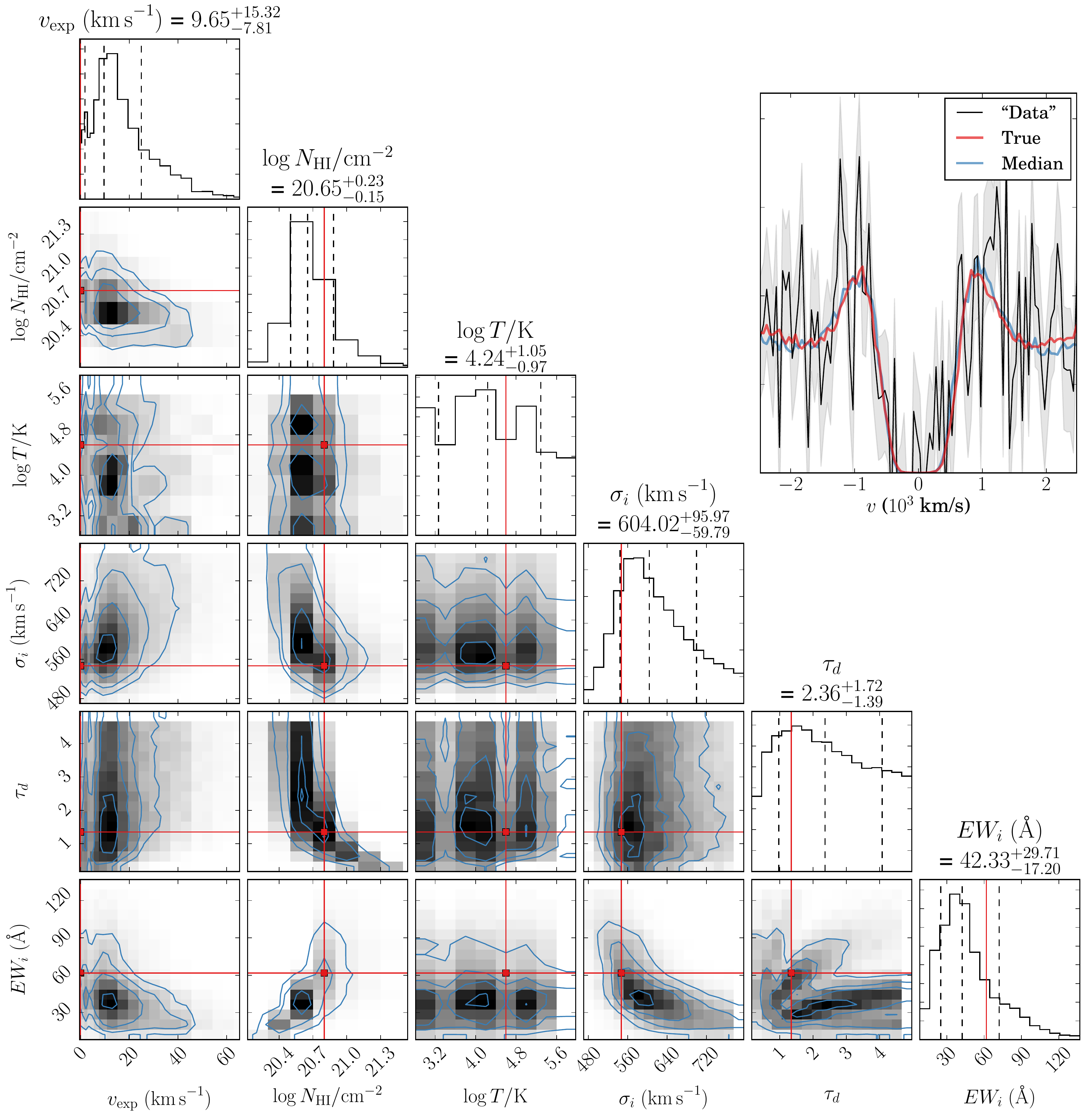}
  \caption{Same as Fig.~\ref{fig:mcmc1}, but for a simulated spectrum with a \Lya absorption feature.}
  \label{fig:mcmc2}
\end{figure*}

\section{Results}
\label{sec:results}

\subsection{Intrinsic degeneracies}
\label{sec:intr-degen}
To investigate whether the best-fit parameters recovered by the standard spectral fitting procedure could be misleading, we performed $640$ local $\Delta^2$-minimizations on each of the $50$ spectra in our fully-simulated observational dataset (see \S~\ref{sec:impl-shell-model}) without adding any additional noise ($n_i = 0$). We chose the number of local minimizations rather arbitrarily, but sufficed to show a few specific cases that highlight the dangers of blindly following this procedure.
We started each minimization from a randomly chosen initial guess. We then discarded the fits where $\Delta^2$ was less than the $\Delta^2$ for the true model parameters. A more systematic study that includes observational uncertainties is presented in \S\ref{sec:uncertainty}.

Fig.~\ref{fig:maxdist_multi} shows examples of degenerate model fits for each of the shell model parameters. In each panel we display the `observed' spectrum (in gray), the spectrum obtained for the true input model parameters (black), and an alternative best-fit model (blue). It can be seen that no shell model parameter is protected from catastrophic recovery errors, as the magnitude of the difference in each of the example cases allows significantly different physical interpretations.

As a concrete example, the central panel in the bottom row of Fig.~\ref{fig:maxdist_multi} shows a spectrum that can be explained with a very high dust content ($\tau_d = 5$), although it was actually produced by a model with very little dust ($\tau_d=0.29$). In fact, the value of $\tau_d=5$ here is the (arbitrarily chosen) upper limit of our parameter space (see \S\ref{sec:impl-shell-model}), so an even higher value may be possible. In this model, the low value of $\sigma_i$, and the high outflow velocity combined with the low hydrogen column density, results in essentially no interaction between the photons and the hydrogen. The path length through the shell (and thus the chance of absorption) is therefore approximately the same for each photon, and so $\tau_d$ has little or no effect on the shape of the resulting spectrum\footnote{In the notation of Eq.~\eqref{eq:photon_weights}, this means that in this particular case $\hat N_{\HI}\sim N_{\HI}$ for \textit{all} photons leading to no spectral shape change due to a change of $\tau_d$.}. The same phenomenon also explains why a $\sim\!1.5$ order of magnitude decrease in \HI column density barely affects the spectrum in the central panel of the top row.

Another case where a parameter hits its prior boundary is in the lower left panel of Fig.~\ref{fig:maxdist_multi}, where $\sigma_i$ reaches its maximum permitted value of $800\kms$. The reason the intrinsic spectral width plays no role in the shape of this absorption feature is because the high hydrogen content causes each \Lya photon to be scattered many times. The relatively large value of $\tau_d$ leads to a high absorption probability, yielding this broad absorption feature. While in this case $\sigma_i$ plays only a minor role, with different values leading to qualitatively similar spectra, the same goodness of fit can only be reached by also changing the other parameters, in particular $\tau_d$ and $EW_i$. This can be explained as follows: while the increase in $\sigma_i$ means that there are initially more \Lya photons in the wing, which have a higher chance of not being absorbed, this is compensated somewhat by the higher dust content. Since this increase also (and more strongly) affects photons travelling longer distances (i.e. the ones closer to the core), a higher number of \Lya photons is also needed to return a similarly good fit as the true model.

The $EW_i\--\tau_d$ degeneracy is also observable in the bottom-right panel, although the difference in $EW_i$ is not as large. This is because, for emission features, $EW_i$ seems to be reasonably well constrained by the continuum level.

Finally, the expansion velocity is thought to be well constrained by the position of the spectral peaks and troughs. The situation shown in the top-left panel is therefore less straightforward -- the effects of changing of $v_{\rm exp}$ can only be compensated by jointly changing $EW_i$, $\sigma_i$, and $\tau_d$.

\begin{figure*}
  \centering
  \plotone{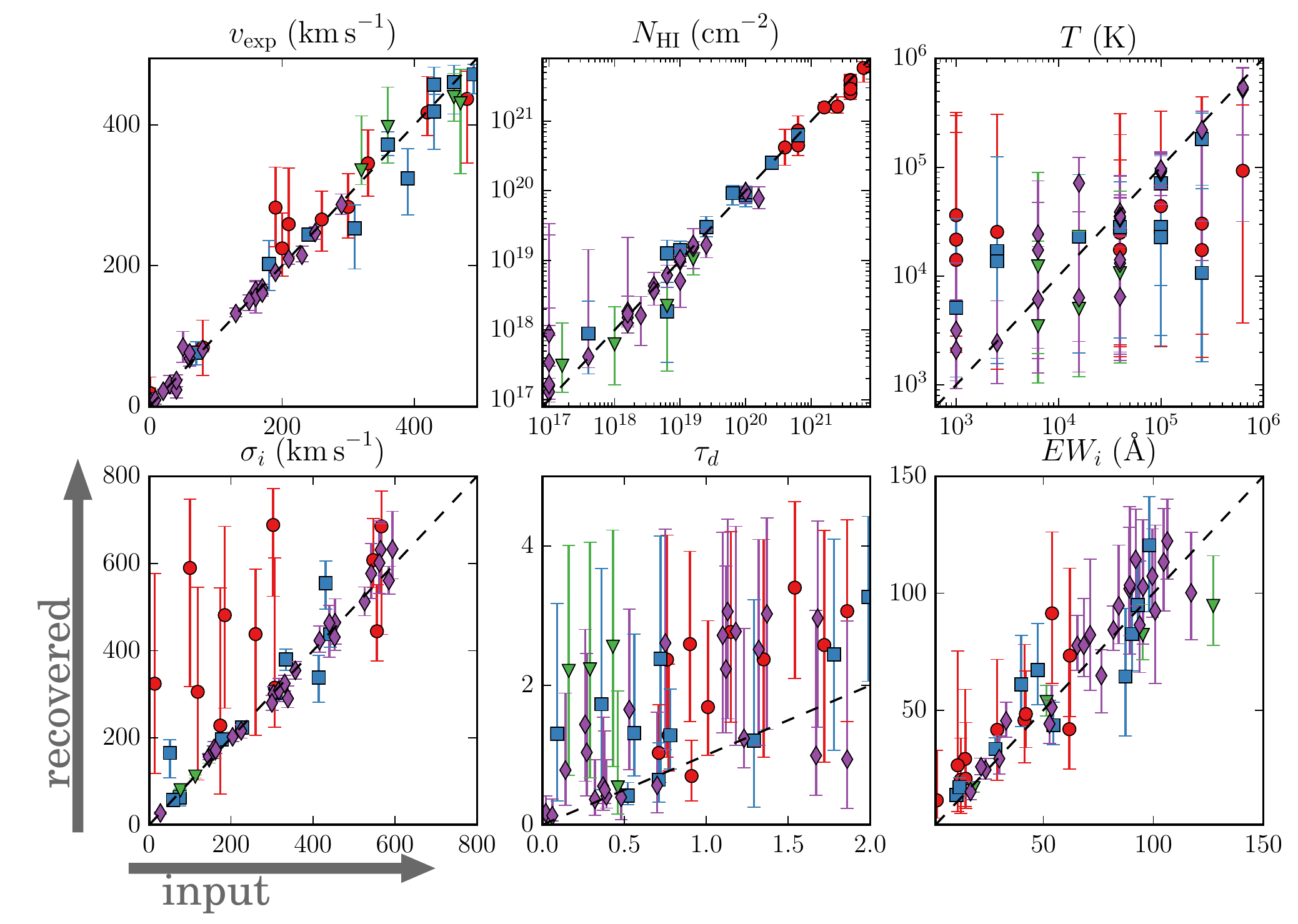}
  \caption{Recovered (y axis) vs. input (x axis) shell model parameters. The error bars mark the 68\% CL, between the $16$th and $84$th percentiles. The colors and symbols indicate different types of spectra: absorption features (red discs), P Cygni-like profiles (blue squares), single-peak (green triangles), and double-peaked (purple diamonds).}
  \label{fig:plot_result_comparison}
\end{figure*}

\subsection{Parameter uncertainty}
\label{sec:uncertainty}
We now turn our attention to the uncertainty associated with the estimated parameters obtained through shell model fitting. First, we consider two example cases, each represented by a spectrum for which the input shell model parameters were correctly recovered. Note that, our simulated dataset contains absorption, single- and double peak emission profiles -- just as observed \Lya spectra \citep[e.g.][]{Steidel2010a,Rivera-thorsen2014,2015arXiv150602885Y}. Out of these, the two cases chosen feature a double peaked emission and a broad absorption profile to show the extent of possible degeneracies. Moreover, in some cases (especially at higher redshift) the blue peak may be further suppressed by the intergalactic medium \citep{2007MNRAS.377.1175D,2011ApJ...728...52L}, in which case our double peaked example is actually more representative of what has been observed.  

We begin by constructing a simulated observation of each spectrum, as described in \S\ref{sec:spectra-fitt-proc}. The noise in each bin is chosen to have an rms of $\hat{\sigma}_i = 0.5\bar I$, where $\bar I$ is the mean intensity of the spectrum. Note that while the noise properties have been chosen rather arbitrarily, the procedure does not depend on this choice, and also works with more realistic data errors. This choice of $\hat{\sigma}_i$ ensures that $\hat{\sigma}_i \gg \hat{\sigma}_{i,{\rm Poisson}}$ (see \S\ref{sec:spectra-fitt-proc}), and can be thought of as representing random errors due to instrumental effects and so on.

This noise level corresponds to signal-to-noise ratios (SNRs) of $\sim 15\--50$, which is comparable with existing surveys \citep[e.g.][]{Adams2011ApJS..192....5A}. We estimated the SNR by maximizing the quantity $\sum_i d_i/\sqrt{\sum_i \hat{\sigma}_i^2}$, where the sums are taken over several adjacent bins. This corresponds to the standard procedure in observing pipelines (S. Wilkins, priv. comm.). For double-peaked profiles and absorption features, different measures are sometimes used -- e.g. taking the difference between the continuum level and the absorption feature -- that tend to result in higher SNR estimates. We will not consider such measures here.
 
Next, we sample the likelihood of Eq.~\eqref{eq:logp} using the affine invariant Markov chain Monte Carlo (MCMC) ensemble sampler \texttt{emcee} \citep{Goodman2010,Foreman-Mackey2012}. We use $900$ walkers with $500$ steps each (including $50$ steps of burn-in). For the starting positions of the walkers, we used the $\Delta^2$-minima found in \S\ref{sec:intr-degen}, weighted by the value of the likelihood at that point, plus a small random perturbation to avoid producing initial paths that are too similar.

Fig.~\ref{fig:mcmc1} shows the results of the MCMC parameter estimation for the first example case\footnote{This and the other triangle plots were produced using a modified version of \texttt{triangle.py} \citep{triangle.py}.}, an asymmetric, double-peaked profile with an estimated SNR of $\sim 32$.
The true (input) shell model parameters are well-recovered, falling within the 68\% credible interval for all but $\tau_d$, which is estimated to be slightly higher than its actual value. One can also see that the expansion velocity and column density are very well constrained (in fact, reaching our grid resolution limit), and that the 1$\sigma$ uncertainty on the temperature is almost half an order of magnitude.

The second example, a spectrum with a broad absorption feature and $\mathrm{SNR} \sim 20$, is shown in Fig.~\ref{fig:mcmc2}. The uncertainties on the estimated shell model parameters are much larger than in the previous example, with all but $N_{\HI}$ being relatively poorly constrained. Degeneracies between many of the parameters can be seen clearly; the values of $\sigma_i$, $EW_i$, $N_\HI$ and $\tau_d$ are all correlated with one another, for the reasons discussed in \S\ref{sec:intr-degen}. Note that the distinct directions visible in the $\tau_d\--EW_i$ plane, and other `striping' features in the plots, are a result of the discrete parametrization of $\log N_{\HI}$, which is mostly localized to the three preferred grid values of $[20.4,\,20.6,\,20.8]$. This was confirmed to be the case by plotting the MCMC steps falling in these three column density bins individually -- it is not an artefact of the sampling procedure. A higher-resolution gridding of the $\log N_{\HI}$ parameter would reduce this effect.\\

Fig.~\ref{fig:plot_result_comparison} shows an overview of the estimated parameters for all $50$ spectra in our simulated dataset, after running exactly the same MCMC procedure as for the two example spectra.
The data points are colored according to spectral type, which was assigned to each spectrum following visual inspection. The parameter estimation procedure reliably recovers at least some of the shell model parameters for all spectral types, particularly the expansion velocity $v_{\rm exp}$, and column density $N_{\HI} \gtrsim 10^{19}\cm^{-2}$ (although the uncertainty is larger for smaller column densities). The intrinsic source parameters $\sigma_i$ and $EW_i$ are also mostly well-recovered, but with a bigger uncertainty. An exception is for spectra with absorption features, where $\sigma_i$ is typically overestimated by $\sim 200\kms$. The worst constraints are obtained for $T$ and $\tau_d$, for which the scatter and uncertainty often reach the prior boundaries for those parameters.

\begin{figure*}
  \centering
  \plotone{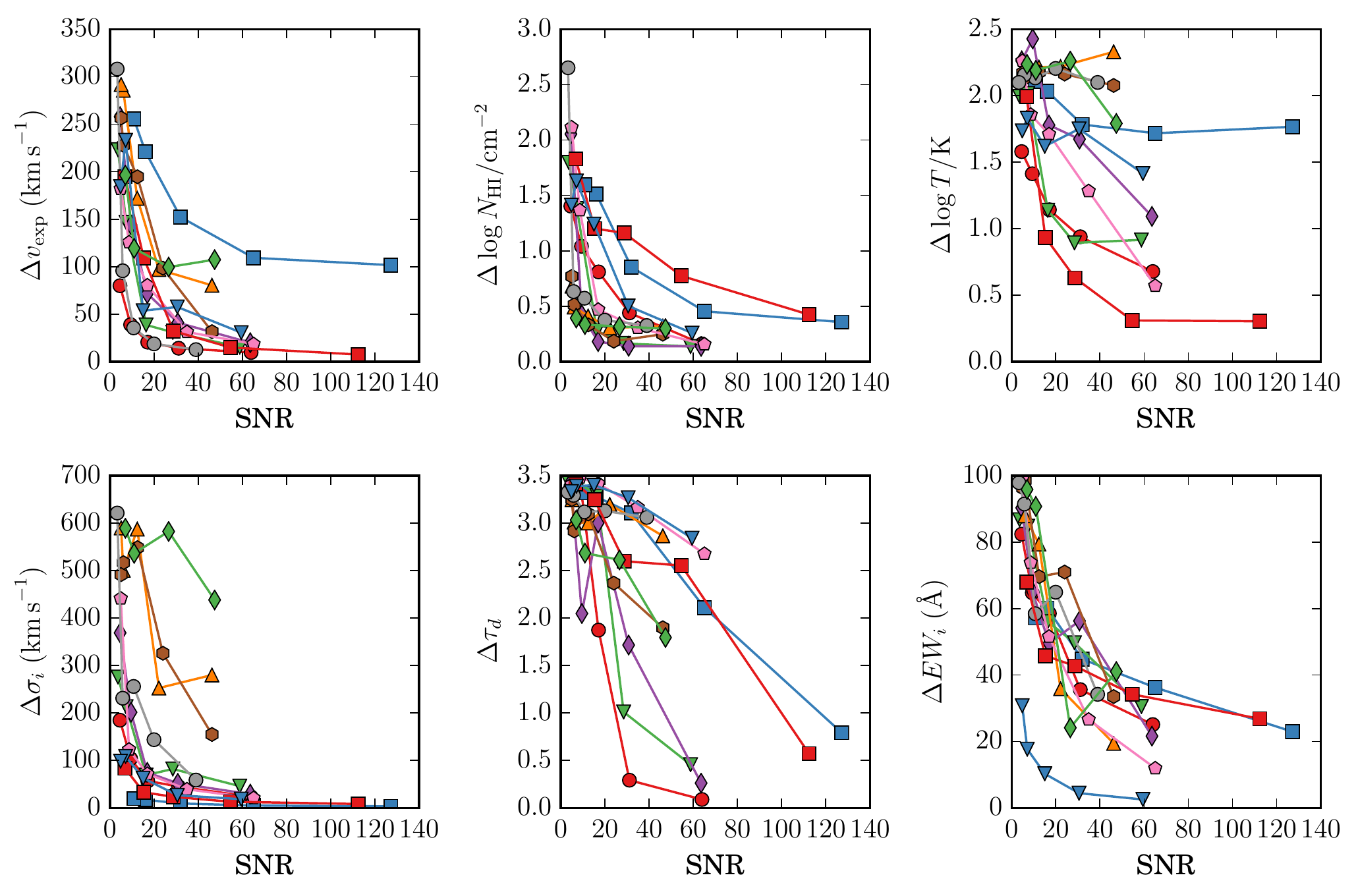}
  \caption{Parameter uncertainties vs. signal-to-noise ratio (SNR) for $12$ different mock spectra, with five choices of $\hat{\sigma}_i$ for each. The y axis shows the 68\% CL error estimate for each parameter, calculated as the difference between the $84$th and $16$th percentiles of the sampled posterior distribution. The SNR is calculated as described in \S\ref{sec:uncertainty}.}
  \label{fig:snr_mixed}
\end{figure*}

\section{Discussion}
\label{sec:discussion}

 Given the large parameter space spanned by the shell model parameters, and the consequent variety of \Lya spectra \citep[as is nicely illustrated in fig.~5 of][]{Schaerer2011}, it is difficult to make foolproof statements about parameter degeneracies and uncertainties. We generally recommend comparing individual spectra to the theoretical predictions on a case-by-case basis instead. In this Section we discuss how we can nevertheless reach more generic conclusions on parameter degeneracies and uncertainties by analysing a sample of $50$ mock spectra.

\subsection{Parameter uncertainties and degeneracies}
\label{sec:param-uncert-degen}

From the results of our likelihood analysis, we find that:

\begin{itemize}
\item \Lya spectra are very sensitive to the expansion velocity, $v_{\rm exp}$. In particular, double-peaked profiles allow the underlying velocity field to be recovered with great accuracy (limited by the simulation grid spacing to $\sim\! 10\kms$ in this study). Absorption features and single peaks lead to greater uncertainties of up to $\sim 50\kms$. This is consistent with the findings of \citet{Verhamme2008}.

\item The intervening column density, $N_{\HI}$, can also be recovered well -- at least for $N_{\HI} \gtrsim 10^{19}{\rm cm}^{-2}$. In this high column density regime, the (68\% CL) uncertainty was limited by our grid spacing to $\Delta\log N_{\HI}/{\rm cm}^{-2}=0.2$, while for lower column densities the uncertainty can be up to one order of magnitude.

\item The effective temperature (including turbulent motion as well as the true temperature) has a complex effect on the \Lya spectrum, and in most cases this parameter cannot be usefully constrained (also see e.g. Verhamme et al. 2008).

\item The width of the intrinsic spectrum, $\sigma_i$, can be recovered for \Lya emission features with very high accuracy ($\Delta\sigma_i\lesssim 50\kms$). For absorption features, on the other hand, this parameter is not constrained so well ($\Delta v_i \sim 200\kms$).

\item We found that the dust content, $\tau_d$, was essentially unconstrained by the shape of the \Lya spectrum. Exceptions to this are for some double-peaked features with low dust content, where the dust content defines the ratio of the heights of the peaks. If, however, measurements of the \Lya escape fraction $f_{\rm esc}$ are available, it is possible to use this extra information in our analysis. When then expect $\tau_d$ to be better constrained in some cases.

\item The intrinsic equivalent width, $EW_i$, can mostly be recovered with an uncertainty of $\sim 20$\AA\ regardless of spectral shape and the value of $EW_i$.
\end{itemize}
Note that these numbers depend on the spectral resolution and quality (SNR). We present the result of changing these quantities in the next section.

We also want to highlight that additional parameters can easily be included into our analysis which might lead to further degeneracies. For example, the line-center of the intrinsic spectrum can be unknown or afflicted with uncertainties when dealing with real data. We found that in some cases this `spectral shift' is degenerate with the outflow velocity $v_{\rm exp}$ and/or the column density $N_{\rm HI}$.

\subsection{Impact of SNR and frequency binning}
\label{sec:snr}
To test the sensitivity of our results to the assumed measurement uncertainties, we randomly selected $10$ out of the $50$ simulated spectra and repeated the analysis of \S\ref{sec:uncertainty} for five different noise levels, $\hat{\sigma}_i/\bar I=(0.25,\,0.5,\,1,\,2,\,4)$. We also include the two spectra analyzed in \S\ref{sec:uncertainty}, for which the resulting SNRs are $\sim\!\!(64,\,32,\,17,\,9,\,4)$ and $(39,\,20,\,10,\,6,\,3)$ respectively. Note that different noise realizations were drawn for each of the noise levels, for each spectrum.

Fig.~\ref{fig:snr_mixed} shows the resulting parameter uncertainties as a function of SNR for each of the $10+2$ spectra.
For $v_{\rm exp}$, $N_{\HI}$, and $\sigma_i$, there is a flattening in the uncertainty beyond $\mathrm{SNR} \sim 20$ for all types of spectrum. A similar behaviour is found for $EW_i$ but at a slightly higher SNR of $\sim 30$, although in this case the curves do not become completely flat, even at the lowest noise levels we considered.
For the remaining two parameters ($\tau_d$, $T$), the evolution of the uncertainty with increasing SNR is not so systematic, with different behaviors for different spectral types. For $T$, some of the spectra show the same flattening at high SNR as the other parameters, for example, while others are considerably more scattered. Note that, because the noise realizations change between the different choices of noise level, it is not surprising that there should be some random variation in the goodness of fit and parameter uncertainties. For $\tau_d$ there is a split between spectra for which the uncertainty rapidly decreases with SNR, and those for which it changes only slightly (presumably in cases where the spectra are relatively independent of $\tau_d$, so it remains poorly constrained).

In summary, for $\mathrm{SNR} \!\sim\! 30$, four of the six shell model parameters can be obtained up to an uncertainty of $(\Delta v_{\rm exp},\,\Delta\log N_\HI / {\rm cm}^{-2},\,\Delta\sigma_i,\,\Delta EW_i)\sim (70\kms,\,0.5,\,100\kms,\,40{\rm \AA})$ for the majority of spectra, and these limits can be further decreased by increasing the SNR.

Finally, we also varied the frequency binning ($\Delta v_i = (25,\,50,\,100,\,200,\,250,\,300,\,500)\kms$) for several of the mock spectra, in order to test the influence of the spectral resolution on the parameter estimates. Figure~\ref{fig:app} in the Appendix shows that our constraints depend weakly on $R$ provided that $R \gsim 10^3$. For lower $R$ the constraints on several parameters (especially $v_{\rm exp}$) increases rapidly for $R < 10^3$, but only for {\it some} models. For example, for the case presented in Fig.~\ref{fig:mcmc1} the characteristic double peak disappears for $\Delta v_i = 250\kms$ ($R\sim 10^3$), which naturally leads to drastically increased uncertainties on many of the parameters. We therefore caution against too general conclusions on the precise impact of $R$, as this strongly depends on the spectrum that is analyzed. 

\section{Conclusion}
\label{sec:conclusion}
The `shell model' for Ly$\alpha$ transfer through galactic outflows has enabled a wide range of \Lya spectra to be characterised in terms of six parameters. An important goal is to understand what physical information is actually encoded in these parameters. This is an important question: {\it if} the parameters of the shell model contain a connection to the underlying physical properties of outflows, then the shell model provides an extremely quick and useful way to capture information about radiative transfer processes on scales that are difficult to model from first principles.\\

As a first step towards addressing this question, we have constructed a large library of shell model spectra and which are available through an interactive online tool. A similar library was presented by \citet{Schaerer2011} which was used in fitting observed \Lya spectra \citep[e.g.,][]{Hashimoto15}. Our work differs in that our grid of spectra is sufficiently closely-meshed\footnote{Quantitatively, we have sampled the $(v_{\rm exp}, \log N_{\HI}, \log T)$ space with 10800 models, while \citet{Schaerer2011} have 780 models to cover this parameter space. In addition, we allowed the dust opacity to vary continuously while \citet{Schaerer2011} ran 8 discrete dust opacities.} to allow a systematic study of possible degeneracies between model parameters, via a fully automated likelihood analysis. In order to make this computationally tractable, we used a post-processing technique to model three of the six shell model parameters. We then applied an automated fitting routine to a small simulated dataset of `noisy' shell model spectra, to see if a likelihood analysis can accurately recover the true (input) model parameters. This presents an important test of the reliability of shell model fitting procedures that are being used in the literature, and provides a simple estimate of the uncertainties on the recovered best-fit model parameters.\\

Our main findings include that from the spectral shape alone it was not possible to provide good constraints on the dust content, $\tau_d$, and effective temperature, $T$, in most models. In contrast, we were able to recover especially the outflow velocity ($v_{\rm exp}$) and the hydrogen column density ($N_{\HI}$) well (especially when $N_{\HI} \gtrsim 10^{19}{\rm cm}^{-2}$). These two results reflect the fact that \Lya radiative transfer observables are most sensitive to the hydrogen distribution and kinematics. Dust affects these observables (emerging spectrum, total flux), but how much depends on other parameters such as $N_\HI$, $v_{\rm exp}$ and $\sigma_i$. This is consistent with observations which indicate that \Lya escape is greatly affected by gas kinematics, and which is likely responsible for the observed large scatter between $f_{esc}$ and $\tau_d$ \citep[see, e.g.,][]{2010ApJ...711..693K,Blanc2011ApJ...736...31B,Hayes2011ApJ...730....8H}.

More quantitatively, for spectra with signal-to-noise ratios $\mathrm{SNR} \!\sim\! 30$ and spectral resolving power $R=6000$, four of the six shell model parameters can be obtained up to an uncertainty of $(\Delta v_{\rm exp},\,\Delta\log N_\HI / {\rm cm}^{-2},\,\Delta\sigma_i,\,\Delta EW_i)\sim (70\kms,\,0.5,\,100\kms,\,40{\rm \AA})$ for the majority of our spectra (see Fig~\ref{fig:snr_mixed} \& \S~\ref{sec:param-uncert-degen}). These constraints can be improved by increasing the SNR (Fig~\ref{fig:snr_mixed}). We found a weak dependence of our results on $R$ down to $R\sim 1000$ (below which our constraints deteriorate faster). However, statements about the impact of the resolving power $R$ depends in detail on the features in the spectrum (e.g. the presence of double peaks clearly can depend on $R$). 

In general, precise quantitative statements on the magnitudes uncertainties and/or the extent of degeneracies depends on the location in the parameter space. (For example, it was not possible to recover $\sigma_i$ specifically for spectra with an absorption feature.) We therefore recommend a case-by-case analysis of observed spectra using a full likelihood analysis (similar to the one presented in \S\ref{sec:uncertainty}). \\

The likelihood analysis allows us to characterise spectra in a systematic, automated way, and produce robust estimates of parameters, their uncertainty, and the degeneracies with other spectra. This procedure will be useful in the near future, when a larger sample of \Lya spectra with sufficient SNR and spectral resolution becomes available. In a follow-up paper (Gronke et al. in prep) we will generate spectra for more realistic models, and fit shell models to these. This analysis will help us take the next \& final step towards understanding what physical information is contained in the shell model parameters. 

\acknowledgments

We thank the anonymous reviewer for useful comments.
M.G. thanks Llu\'{i}s Mas-Ribas and Stephen Wilkins for helpful advice and useful discussions and Michael Rauch for inspiring the online tool during EWASS $2014$. P.B. is supported by European Research Council grant StG2010-257080.

\bibliography{references_all}

\appendix
\section{Varying the Resolving Power $R$}
\label{sec:varying_R}
For the main results we presented in the paper we assumed $R=6000$. Here, we quantify the impact of varying the spectral resolution/resolving power while keeping the SNR fixed at $\sim 35$. Note that  we generate a different noisy random realisation of a given shell model for each $R$. This explains why our constraints do not vary monotonically with $R$ for a given model. 

\begin{figure*}
  \centering
  \plotone{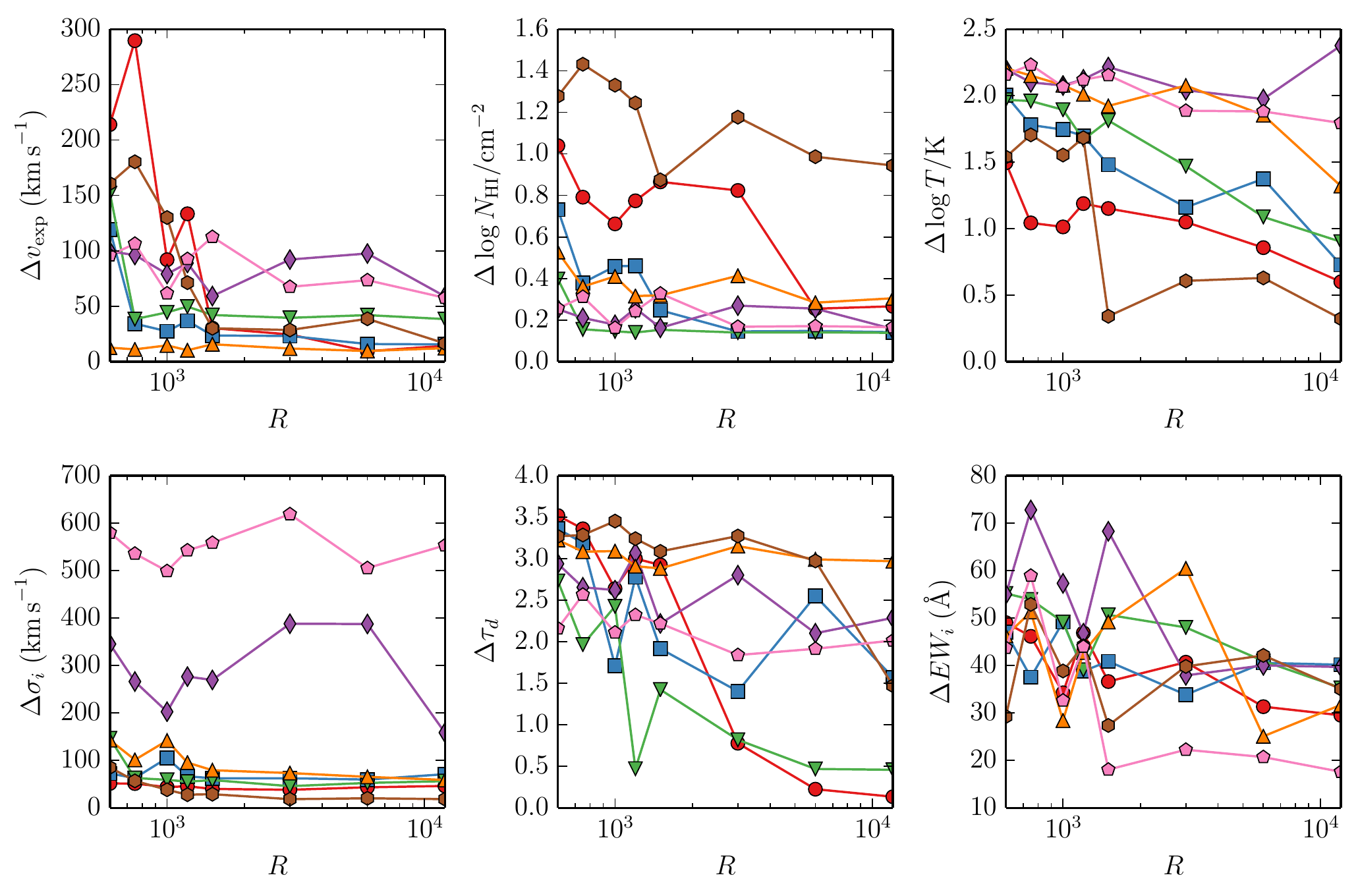}
  \caption{Same as \ref{fig:snr_mixed} but here we varied the resolving power $R$ (see Appendix~\ref{sec:varying_R}). Our inferred constraints generally depend weakly on $R$ for $ R\gsim 10^3$. For lower resolving power our constraints start to deteriorate faster.}
  \label{fig:app}
\end{figure*}

\end{document}